\documentclass[conference]{IEEEtran}
\usepackage{amsmath}
\usepackage{amsfonts}
\usepackage{amssymb}
\usepackage{amsbsy}
\usepackage{graphicx}
\usepackage{times}
\usepackage{rotating}
\usepackage{bm}
\usepackage{bbm}
\usepackage{color}
\usepackage{cite}
\usepackage{enumerate} 
\usepackage{dsfont}
\usepackage{multirow}
\usepackage{subfigure}

\newtheorem{theorem}{Theorem}

\newtheorem{lemma}{Lemma}

\newcommand{\be}{\begin{equation}}
\newcommand{\ee}{\end{equation}}
\newcommand{\ist}{\hspace*{.3mm}}
\newcommand{\rmv}{\hspace*{-.3mm}}
\providecommand{\abs}[1]{\lvert#1\rvert}

\providecommand{\norm}[1]{\lVert#1\rVert}

\def\L{N}

\def\R{R}
\def\Q{Q}

\def\T{T}

\def\0v{\boldsymbol{0}}
\def\Iv{\text{\bf I}}

\def\sv{\boldsymbol{s}}
\def\svt{\tilde{\sv}}

\def\uv{\boldsymbol{u}}

\def\vv{\boldsymbol{v}}

\def\xv{\boldsymbol{x}}
\def\xvt{\tilde{\xv}}
\def\yv{\boldsymbol{y}}
\def\yvb{\bar{\yv} }  

\def\zv{\boldsymbol{z}}

\def\xiv{\boldsymbol{\xi}}

\def\Am{\boldsymbol{A}}

\def\Bm{\boldsymbol{B}}

\def\Jm{\boldsymbol{J}}

\def\Qm{\boldsymbol{z}}
\def\Qmt{\tilde{\Qm}}

\def\Xim{\rBm}

\def\Zm{\boldsymbol{Z}}

\def\Sigm{\boldsymbol{\Sigma}}

\def\rs{\mathsf{s}}

\def\randk{\mathsf{k}}

\def\r0v{\boldsymbol{\mathsf{0}}}

\def\rhv{\boldsymbol{\mathsf{h}}}

\def\rnv{\boldsymbol{\mathsf{w}}}
\def\rnvt{\rnv'}

\def\rsv{\boldsymbol{\mathsf{s}}}

\def\ruv{\boldsymbol{\mathsf{u}}}

\def\rvv{\boldsymbol{\mathsf{v}}}

\def\rxv{\boldsymbol{\mathsf{x}}}

\def\ryv{\boldsymbol{\mathsf{y}}}
\def\ryvb{\bar{\ryv} }  
\def\ryvt{\tilde{\ryv}}

\def\rAm{\boldsymbol{\mathsf{A}}}

\def\rBm{\boldsymbol{\mathsf{B}}}

\def\IN{\mathbb{N}}
\def\IC{\mathbb{C}}
\def\IR{\mathbb{R}}
\def\IZ{\mathbb{Z}}

\def\sD{\mathcal{D}}

\def\sI{\mathcal{I}}

\def\sM{\mathcal{M}}
\def\sN{\mathcal{N}}

\def\sP{\mathcal{P}}

\def\sU{\mathcal{U}}

\providecommand{\absdet}[1]{\lvert#1\rvert}
\def\diag{\operatorname{diag}}

\def\trans{{\operatorname{T}}}
\def\E{\mathbb{E}}

\begin{document}

\IEEEoverridecommandlockouts

\allowdisplaybreaks

\title{
Generic Correlation Increases  \\ Noncoherent MIMO Capacity
}

\author{
\IEEEauthorblockN{G\"unther Koliander$^1$, Erwin Riegler$^1$, Giuseppe Durisi$^2$, and Franz~Hlawatsch$^1$}\\
\IEEEauthorblockA{
$^1$Institute of Telecommunications, Vienna University of Technology, 1040 Vienna, Austria\\
$^2$Department of Signals and Systems, Chalmers University of Technology, 41296 Gothenburg, Sweden
}
\thanks{\hspace*{-2.7mm}This work was supported by the WWTF under grant ICT10-066 (NOWIRE).}
}                                                                                               

\maketitle

\begin{abstract}
We study
the high-SNR capacity of MIMO Rayleigh block-fading channels in the noncoherent setting where neither transmitter nor receiver has \emph{a priori} channel state information. 
We show that when the number of receive antennas is sufficiently large and 
the temporal correlation within each block is ``generic'' (in the sense used in the interference-alignment literature), the capacity 
pre-log is given by $\T(1-1/\L)$ for $\T<\L$, where $\T$ denotes the number of transmit antennas and $\L$ denotes the block length.
A comparison with the 
widely used constant block-fading channel (where the fading 
is constant within each block)
shows that for a large 
block length, 
generic correlation increases the capacity pre-log by a factor of about four.
\vspace{1mm}
\end{abstract}

\section{Introduction} \label{sec:intro}   

The throughput achievable with 
multiple-input multiple-out\-put (MIMO) wireless systems is 
limited by the need to acquire channel state information (CSI)~\cite{adca12}.
A fundamental way to assess the corresponding rate penalty 
is to study capacity in the 
\emph{noncoherent setting} where neither the transmitter nor the receiver has \emph{a priori} CSI.
 
We consider a MIMO system with $T$ transmit antennas and $R$ receive antennas. In the widely used 
\emph{constant block-fading channel model} 
\cite{MaHo99},
the fading process takes on independent realizations across blocks of  $\L$ channel uses (``block-memoryless'' assumption), 
and within each block the fading coefficients are constant. 
Thus,
the $\L$-dimensional channel gain vector  describing 
the channel between antennas $t$ and 
$r$ 
(hereafter briefly termed ``$(t,r)$ channel'') within a block 
is 
%
%
\begin{equation}\label{eq:constant_block_memoryless}
  \rhv_{r,t}=\rs_{r,t} \ist \mathbf{1}_{\L} \,.
\end{equation}
Here, $\mathbf{1}_{\L}$ denotes the $\L$-dimensional all-one vector and $\{\rs_{r,t}\}_{r \in \{1,\dots,R\},\ist t \in \{1,\dots,T\}}$ 
are independent $\mathcal{CN}(0,1)$ random variables.
Unfortunately, even for this simple channel model, a closed-form expression of noncoherent capacity 
is unavailable.
However,
an accurate characterization exists for
high signal-to-noise ratio (SNR) values.
In~\cite{zhengtse02}, it was shown that the capacity pre-log (i.e., the asymptotic ratio between capacity and the logarithm of the SNR as the SNR grows large) 
for the constant block-fading model is given by 
\begin{equation}
\hspace*{-.5mm}\chi_{\text{const}}=\ist M \bigg(\rmv 1 \rmv-\frac{M}{\L} \bigg) \ist , \;\;\, \text{with} \;\, M \rmv = \ist \min\{\T, \R, \lfloor\L/2\rfloor\} \ist .\label{eq:pre_log_constant}
\end{equation}
%
A more detailed high-SNR capacity expansion 
was obtained in \cite{zhengtse02} for the case $\R+\T\leq \L$; this expansion was recently extended
in \cite{yaduri12} to the large-MIMO setting $\R+\T>\L$.


One limitation
of the constant block-fading model is that it fails to describe 
a specific setting where block-fading models are of interest, namely, cyclic-prefix orthogonal frequency division multiplexing (CP-OFDM) systems \cite{tsvi05}.
In such systems, the channel input-output relation is most conveniently described in the frequency domain; the 
vector of channel gains  $\rhv_{r,t}$ is then equal to the Fourier transform of the discrete-time impulse response of
the $(t,r)$ channel.
Let us assume that  $\rhv_{r,t}$ changes independently across blocks of length $\L$ and that
\begin{align}\label{eq:generic_block_memoryless}
  \rhv_{r,t} = \rs_{r,t} \ist \zv_{r,t} \,,
\end{align}
where $\zv_{r,t}$ is a deterministic vector whose squared inverse Fourier transform equals
the power-delay profile of the $(t,r)$ channel and, as before, $\{\rs_{r,t}\}_{r \in \{1,\dots,R\},\ist t \in \{1,\dots,T\}}$ 
are independent $\mathcal{CN}(0,1)$ random variables.
As the vectors $\zv_{r,t}$ are related to power-delay profiles, it is reasonable to assume that they are different for different $(t,r)$.
Note that the constant block-fading model \eqref{eq:constant_block_memoryless} is a special case of \eqref{eq:generic_block_memoryless} in which the  
impulse response of each $(t,r)$ channel consists of only a single tap, a case for which the use of OFDM is unnecessary.


\subsubsection*{Contributions} 
We study
the capacity pre-log 
(hereafter briefly termed ``pre-log'')
of 
MIMO block-fading channels modeled as in
\eqref{eq:generic_block_memoryless}. We show that when the deterministic vectors $\{\zv_{r,t}\}$ are 
\emph{generic},\footnote{We 
use the term ``generic''  in the same sense as in the interference-alignment literature \cite{ja11}. 
A rigorous definition will be provided in Section \nolinebreak 
\ref{sec:syst}.}Ê
the pre-log can be larger than the pre-log 
in the constant block-fading case as given in~\eqref{eq:pre_log_constant}.
Specifically, we show that for the \emph{generic block-fading model} (i.e., the 
model \eqref{eq:generic_block_memoryless}
with generic vectors $\{\zv_{r,t}\}$), when $T \!\!<\!\rmv \L$ and the number of receive antennas is sufficiently large such that $R\geq {T(\L \rmv-\rmv 1)/(\L \rmv-\rmv  T)}$, the 
pre-log is given by
\be\label{eq:pre_log_generic}
  \chi_{\text{gen}}=T\bigg(1\rmv-\rmv \frac{1}{\L}\bigg) \ist.
\ee
%
For large $\L$, the highest achievable $\chi_{\text{gen}}$ (with appropriately chosen $T$ and $R$)
is about four times as large as the highest achievable $\chi_{\text{const}}$.
As we will demonstrate, this is because under the generic block-fading model,
the received signal vectors 
in the absence of noise span a subspace of higher dimension than under the constant block-fading model. 

To establish~\eqref{eq:pre_log_generic}, we derive
an upper bound on the pre-log of the model \eqref{eq:generic_block_memoryless}.
This upper bound matches asymptotically
the pre-log lower bound that was recently developed in~\cite{koridumohl12} in a more general setting 
(the generic block-fading model considered in this paper is a special case of the system model in \cite{koridumohl12} for correlation rank $\Q \!=\! 1$).
Thus, the combination of the two bounds establishes the pre-log expression~\eqref{eq:pre_log_generic}. 
As the proof in \cite{koridumohl12} is rather involved, we also illustrate the main ideas of the proof of the lower bound using an example. 
In this illustration, we present a new method for bounding the change in differential entropy that occurs when a random variable undergoes a finite-to-one mapping; this method significantly simplifies one step in the proof. 



\subsubsection*{Notation}
Sets are denoted by calligraphic letters (e.g., $\sI$), and $|\mathcal{I}|$ 
denotes the cardinality of $\mathcal{I}$. The indicator function of a set $\sI$ is denoted by $\mathbbmss{1}_{\sI}$. 
We use the notation $[M\!:\!N]\triangleq$\linebreak 
$\{M, M\!+\rmv 1,\dots,N\}$ for $M, N \!\in\! \IN$.
Boldface uppercase (lower\-case) letters denote matrices (vectors). Sans serif letters denote random quantities,  e.g., $\rAm$ is a random matrix, 
$\rxv$ is a random vector, and $\rs$ is a random scalar. 
The superscripts ${}^{\operatorname{T}}$ and ${}^{\operatorname{H}}$ stand for transposition and 
Hermitian transposition, respectively. 
The all-zero vector of appropriate size is written as $\0v$, and the $M \rmv\times\rmv M$ identity matrix as
$\Iv_{M}$. The entry
in the $i$th row and $j$th column of a matrix $\Am$
is denoted by $[\Am{]}_{i,j}$, and the $i$th entry of a vector $\xv$ by $[\xv{]}_i$.
We denote by $\diag(\xv)$ the diagonal matrix with the entries of $\xv$ in its main diagonal,
and by $\abs{\Am}$ the modulus of the determinant of a square matrix $\Am$.
For $x \!\in\!\IR$, we define $\lfloor x\rfloor\triangleq \max\{m \!\in\! \IZ \!\mid\! m \!\leq\! x\}$. 
We write $\E[\cdot]$ for the expectation operator,
and $\rxv\sim\mathcal{CN}(\0v,\Sigm)$ to indicate
that $\rxv$  is a circularly symmetric complex Gaussian random vector with covariance matrix $\Sigm$.
The Jacobian matrix of a differentiable function $\phi$ is denoted by $\Jm_{\phi}$.

\vspace{.7mm}

\section{System Model} \label{sec:syst}   

For
the 
block-fading channel defined by \eqref{eq:generic_block_memoryless}, the input-out\-put relation for a given block of length~$\L$ 
\vspace{-.3mm}
is 
\be\label{eq:model1}
\ryv_{r} \ist=\, \sqrt{\frac{\rho}{\T}} \!\sum_{t\in [1:\T]} \! \rs_{r,t} \ist \Zm_{r,t} \ist \rxv_{t} \ist+\, \rnv_{r} \,,\quad r\in [1\!:\!\R] \,.
\vspace{-.8mm}
\ee
Here,
$\rxv_{t} \!\in\! \IC^{\L}$ is the signal vector transmitted by the $t$th \nolinebreak 
trans\-mit \nolinebreak 
antenna; 
$\ryv_{r} \!\in\! \IC^{\L}$ is the vector received by the $r$th receive antenna; 
$\rs_{r,t} \!\sim\! \mathcal{CN}(0,1)$ is a random variable describing the $(t,r)$ channel;
$\Zm_{r,t}\triangleq \diag(\Qm_{r,t})$, where $\zv_{r,t}$ is a deterministic vector; 
$\rnv_{r} \!\sim \mathcal{CN}(\0v,\Iv_{\L})$ is the noise vector at the $r$th receive antenna; 
and $\rho \!\in\! \IR^+$ is the SNR. 
If 
$\Zm_{r,t} \rmv=\rmv \Iv_{\L}$ for all $r\in [1\!:\!\R]$ and $t\in [1\!:\!\T]$, then \eqref{eq:model1} reduces to the constant block-fading model.
We assume that all $\rs_{r,t}$ and $\rnv_{r}$ are mutually independent
and independent 
across different blocks, and that
the 
vectors $\rxv_{t}$ are 
independent of 
all $\rs_{r,t}$ and $\rnv_{r}$.

For later use, we define the vectors
$\rxv\triangleq (\rxv_{1}^{\operatorname{T}} \cdots\ist \rxv_{\T}^{\operatorname{T}})^{\operatorname{T}}\!\in\IC^{\T\L}\rmv$,
$\ryv \triangleq (\ryv_{1}^{\operatorname{T}} \cdots\ist \ryv_{\R}^{\operatorname{T}})^{\operatorname{T}}\!\in\rmv\IC^{\R\L}\rmv$, and 
$\rnv \triangleq (\rnv_{1}^{\operatorname{T}} \cdots\ist \rnv_{\R}^{\operatorname{T}})^{\operatorname{T}}\!\in\IC^{\R\L}$
and the matrix $\Zm \triangleq ( \Qm_{r,t} {)}_{r\in [1:\R], \ist t\in [1:T]} \!\in\rmv \IC^{\R\L\times \T}\rmv$.
We will 
use the phrase ``for a generic correlation'' or ``for a generic $\Zm$'' to indicate
that a property holds for almost every matrix $\Zm$, which means more specifically that the set of all $\Zm$ for which the 
property does \emph{not} hold has Lebesgue measure zero.


\section{Pre-log Characterization} \label{sec:q1}   

\vspace{1mm}

\subsection{Main Result} 
\label{sec:main_result}

Because of the block-memoryless assumption, the coding theorem in \cite[Section 7.3]{Gallager68} implies that the capacity of the channel~\eqref{eq:model1} is given 
\vspace{-.4mm}
by 
\be\label{eq:capacity}
C(\rho) \ist=\ist \frac{1}{\L}\sup  I(\rxv \ist;\ryv) \,.
\ee
Here, $I(\rxv \ist;\ryv)$ denotes mutual information \cite[p.\,251]{Cover91} and the supremum is taken over all input distributions on $\IC^{\T\L}$
that satisfy the average power constraint
\[
\E[\|\rxv\|^2] \ist\leq\ist \T\L \,.
\]
The pre-log is then defined 
\vspace{-.7mm}
as  
\be\label{eq:prelog1}
\chi \ist \triangleq\lim_{\rho\to\infty}\frac{C(\rho)}{\log(\rho)} \,.
\ee

Our main result is the following 
\vspace{1.5mm}
theorem.

\begin{theorem}\label{THmaintheoremq1}
Let $\T \!<\rmv \L$ and $\R\geq \T(\L \rmv-\rmv 1)/(\L \rmv-\rmv \T)$. For a generic correlation, the pre-log of the channel \eqref{eq:model1} is 
given by \eqref{eq:pre_log_generic}, i.e.,
$\chi_{\text{gen}}=\ist
\T(1\rmv-\rmv 1/\L)$.
\vspace{1.5mm}
\end{theorem} 

\begin{IEEEproof}[\hspace{-1em}Proof]
In Section~\ref{sec:upperbound}, we will 
show that the pre-log is upper-bounded by $\T\ist (1\rmv-\rmv 1/\L)$.
For $\T \!<\rmv \L$, $\R\geq \T(\L \rmv-\rmv 1)/(\L \rmv-\rmv \T)$, and a generic correlation,
this pre-log is achievable as a consequence of the lower bound in \cite[Theorem~1]{koridumohl12}.
\end{IEEEproof}

\vspace{1mm}

\subsection{Pre-log Gain}
\label{sec:rationale_behind_the_pre_log_gain}

For the constant block-fading model \eqref{eq:constant_block_memoryless}, it follows from \eqref{eq:pre_log_constant} that the pre-log 
is maximized for $T = R = \lfloor\L/2\rfloor$, which yields $\chi_{\text{const}}=\lfloor\L^2/2\rfloor/(2\L)\leq \L/4$.
In contrast, for the generic block-fading model \eqref{eq:generic_block_memoryless} with $\T \rmv<\rmv \L$, it follows from 
\eqref{eq:pre_log_generic} that the pre-log is maximized for $\T=\L \rmv-\rmv 1$ and $\R=(\L \rmv-\rmv 1)^2$, which results in 
$\chi_{\text{gen}}=(\L \rmv- 1)^2/\L$. 
For large $\L$, this is about four times as large as the highest achievable $\chi_{\text{const}}$.
We will now provide some intuition regarding
this pre-log gain. For concreteness and simplicity, we 
consider the case $\T \rmv=\rmv 2, \R \rmv=\rmv 3, \L \rmv=\rmv 4$.

The pre-log can be interpreted 
as the 
number of entries of $\rxv \rmv\in\rmv \IC^{8}$ that can be deduced from a received $\ryv \rmv\in\rmv \IC^{12}$ 
in the absence of noise, 
divided by the block length (coherence 
length) $\L \!\rmv=\! 4$.
In the constant block-fading model, the 
noise\-less 
re\-ceived vectors $\ryvb_r \rmv= \rs_{r,1}\rxv_1 \rmv+ \rs_{r,2} \ist \rxv_2$, $r \rmv=\rmv 1,2,3$ 
belong to the two-dimensional subspace spanned by $\{\rxv_1, \rxv_2\}$. 
Hence, the received vectors $\ryvb_1, \ryvb_2, \ryvb_3$ are linearly dependent, and any two of them 
contain all the information available about $\rxv$. 
From, e.g., 
$\ryvb_1$ and $\ryvb_2$, we obtain $2\cdot4$
equations in the $8 + 4$
variables ($\rxv,
\rs_{1,1}, \rs_{1,2}, \rs_{2,1}, \rs_{2,2}$). Since we do not have control\linebreak 
of the variables $\rs_{r,t}$, one way to reconstruct $\rxv$ is to fix
four
of its entries (or, equivalently, to transmit four
pilot symbols) to obtain
eight
equations in eight
variables. By solving this system of equations, we obtain four
entries of $\rxv$, which corresponds to
a pre-log of $4/4=\rmv 1$.

In the generic block-fading model, on the other hand, the noiseless received vectors 
$\ryvb_r \rmv= \rs_{r,1}\Zm_{r,1} \ist \rxv_1 + \rs_{r,2}\Zm_{r,2} \ist \rxv_2$, $r \rmv= 1,$ $2,3$ can span a three-dimensional subspace. 
Hence, we obtain a system of $3\cdot 4$ equations in the $8 + 6$ variables ($\rxv,
\rs_{1,1}, \rs_{1,2}$, $\rs_{2,1}, \rs_{2,2}, \rs_{3,1}, \rs_{3,2}$). 
Fixing
two
entries of $\rxv$, we are able to recover the remaining six
entries.
Hence, the pre-log is $6/4=3/2$.
These arguments suggest that the reason why the generic block-fading
model yields a larger pre-log than the constant block-fading model is that 
the noiseless received vectors span a subspace of $\IC^{\L}$ of higher dimension.

\vspace{.5mm}

\section{Upper bound} \label{sec:upperbound}   

The following upper bound on the pre-log of the 
channel \eqref{eq:model1} 
holds for arbitrary 
$\T$, $\R$, $\L$, and  
\vspace{1mm}
$\Zm$.

\begin{theorem}\label{THupperbound}
The pre-log of the channel \eqref{eq:model1} satisfies 
\be\label{eq:upperbound}
\chi \ist\leq\ist \T\bigg(\rmv 1 \rmv-\rmv \frac{1}{\L}\rmv \bigg)\ist .
\ee
\end{theorem}

\begin{IEEEproof}[\hspace{-1em}Proof]
We will show that the 
pre-log is upper-bounded by $\T$\linebreak 
times the pre-log of a constant block-fading single-input mul\-tiple-output (SIMO) channel. The 
result then follows from \eqref{eq:pre_log_constant}.

From \eqref{eq:model1},
the input-output relation at time
$n \!\in\! [1\!:\!\L]$ is
\be
[\ryv_r{]}_n =\ist \sqrt{\frac{\rho}{\T}} \!\rmv \sum_{t\in[1:\T]} \!\!  \rs_{r,t} \ist  [\Qm_{r,t}{]}_{n} \ist [\rxv_t{]}_n +\ist [\rnv_r{]}_n\,, \quad\!  r \!\in\! [1\!:\!\R]\,. \;
\label{eq:singleinout}
\ee
Consider now $\T$ constant block-fading SIMO channels with $\R$ receive antennas and SNR equal to $K\rmv\rho$, where $K$ is any finite constant satisfying
$K \!>\rmv \max_{r\in [1:\R],\ist n\in[1:\L]}\sum_{t\in[1:\T]}\abs{[\Qm_{r,t}{]}_{n}}^2$. The input-output relation of the $t\ist$th SIMO channel, 
with $t \!\in\! [1\!:\!\T]$, 
is
\be\label{eq:SIMO}
[\ryvt_{r,t}{]}_n =\ist \sqrt{K\rho} \,\ist \rs_{r,t} \ist [\rxv_t{]}_n +\ist [\tilde\rnv_{r,t}{]}_n \,,\quad\!  r \!\in\! [1\!:\!\R]\,.
\ee
We can rewrite \eqref{eq:singleinout} using \eqref{eq:SIMO} as follows: 
\be\label{eq:MIMO_SIMO}
[\ryv_r{]}_n =\ist \frac{1}{\sqrt{K\T}} \!\sum_{t\in[1:\T]} \! [\Qm_{r,t}{]}_{n} \ist [\ryvt_{r,t}{]}_n +\ist [\rnvt_r{]}_n \,,
\vspace{-.3mm}
\ee
where the $[\rnvt_r{]}_n \rmv\sim [\rnv_r{]}_n \rmv- \sum_{t\in [1:\T]}[\Qm_{r,t}{]}_{n}\ist [\tilde\rnv_{r,t}{]}_n/\sqrt{K\T} 
\rmv\sim \mathcal{CN} \big( 0, 1 - \sum_{t\in[1:\T]} \abs{[\Qm_{r,t}{]}_{n}}^2/(K\T) \big)$ 
are mutually independent and independent of all $\rxv_t, \rs_{r\!,t}$, and $\tilde\rnv_{r\!,t}$. 
The additional noise terms $[\rnvt_r{]}_n$ ensure that the total noise in \eqref{eq:MIMO_SIMO} has unit variance. 
The data-processing inequality applied to \eqref{eq:MIMO_SIMO} 
yields
\be\label{eq:dataproc1}
I(\rxv\ist ;\ryv) \ist\leq\ist I(\rxv\ist ;\ryvt_1,\dots, \ryvt_{\T}) \,,
\vspace{.3mm}
\ee
with $\ryvt_t \triangleq (\ryvt_{1,t}^{\operatorname{T}} \cdots\ist \ryvt_{\R,t}^{\operatorname{T}})^{\operatorname{T}}\!\in\rmv\IC^{\R\L}\rmv$.
The right-hand side of \eqref{eq:dataproc1} can be upper-bounded as follows:
\begin{align}
I(\rxv\ist ;\ryvt_1,\dots, \ryvt_{\T}) &\,=\, h(\ryvt_1,\dots, \ryvt_{\T}) \ist-\ist h(\ryvt_1,\dots, \ryvt_{\T} |\ist \rxv) \notag \\[.5mm]
& \stackrel{(a)}= h(\ryvt_1,\dots, \ryvt_{\T}) \ist-\! \sum_{t\in[1:\T]} \!h(\ryvt_t | \ist \rxv_t) \notag \\[-.3mm]
& \ist\ist\leq \!\sum_{t\in[1:\T]} \!\! \big[ h(\ryvt_t) - h(\ryvt_t | \ist \rxv_t) \big] \notag \\[.5mm]
& \,=\sum_{t\in[1:\T]} \!\! I(\rxv_t\ist ;\ryvt_t) \notag \\[-1mm]
& \ist\stackrel{(b)}\leq \T\L \ist C_{\text{const}}(K\rho) \notag \\[0mm]
& \ist\stackrel{(c)}=\ist \T(\L \!-\! 1)\log(K\rmv\rho) \ist+\ist o(\log(\rho)) \notag \\[.5mm]
& \,=\, \T(\L \!-\! 1)\log(\rho) \ist+\ist o(\log(\rho))\,. 
\label{eq:upperbound1}
\end{align}
Here, $h$ denotes differential entropy, 
$(a)$ holds because $\ryvt_1,\dots,$\linebreak 
$\ryvt_{\T}$ are conditionally independent given $\rxv$, 
$(b)$ follows from \eqref{eq:capacity} (note that $C_{\text{const}}(K\rho)$ refers to the capacity of constant block-fading SIMO channels), 
and $(c)$ follows from \eqref{eq:prelog1} and \eqref{eq:pre_log_constant} for $M=1$.
Inserting \eqref{eq:upperbound1} into \eqref{eq:dataproc1} and using \eqref{eq:capacity} yields
\[
C(\rho) \ist\leq\ist \T\ist\frac{\L \!-\! 1}{\L}\log(\rho) \ist+\ist o(\log(\rho))\,,
\]
from which \eqref{eq:upperbound} follows via \eqref{eq:prelog1}.
\end{IEEEproof}

\vspace{.8mm}

\section{Lower bound} \label{sec:lowerbound}   

According to \cite[Theorem~1]{koridumohl12}, for $\T \!<\rmv \L$ and $\R\geq \T(\L \rmv-$\linebreak 
$1)/(\L \rmv-\rmv \T)$,
the pre-log of the generic block-fading channel \eqref{eq:model1} is lower-bounded by
$\chi_{\text{gen}} \ge \T \ist (1\rmv-\rmv 1/\L)$.
We will now illustrate the main ideas of the proof of this lower bound and present a new method for bounding the change in differential entropy under 
a finite-to-one mapping (Lemma~\ref{LEMchangeh} in Section~\ref{sec:proof}), which significantly simplifies one of the steps of the proof. 
For concreteness, we consider the special choice $\T \!=\rmv 2$, $\R \rmv=\rmv 3$, and $\L \rmv=\rmv 4$. For this choice, $\T\ist (1\rmv-\rmv 1/\L) = 3/2$.

In the remainder of this paper, we choose the input distribution $\rxv \sim \mathcal{CN}(\0v,\Iv_{8})$. 
Because of \eqref{eq:capacity} 
and \eqref{eq:prelog1}, we obtain
\be \label{eq:preloggauss}
\chi \, \geq\ist \frac{1}{4} \lim_{\rho\to\infty}\frac{I(\rxv \ist;\ryv)}{\log(\rho)} \,.
\vspace{-2mm}
\ee
Since
\vspace{.5mm}
\be\label{eq:sepixy}
I(\rxv \ist;\ryv) \ist=\ist h(\ryv)-h(\ryv\ist|\ist\rxv)\,,
\ee
we can lower-bound $I(\rxv \ist;\ryv)$ by lower-bounding $h(\ryv)$ and
upper-bounding $h(\ryv\ist |\ist\rxv)$. 
For later use, we note that the input-output relation \eqref{eq:model1} can be written as
\be\label{eq:modelsimp}
\ryv\,=\ist \sqrt{\frac{\rho}{2}} \ist \ryvb \ist+\ist \rnv \,,
\vspace{-2mm}
\ee
with
\vspace*{-2.5mm}
\be\label{eq:ybarsimp}
\ryvb \,\triangleq\ist
\underbrace{\begin{pmatrix}
\begin{matrix}
\Zm_{1,1}\rxv_1 \hspace*{-2mm} &\hspace*{-4mm}&\hspace*{-2mm} \\[-.3mm]
\hspace*{-2mm} & \hspace*{-2mm} \Zm_{2,1}\rxv_1 \hspace*{-2mm} \\[-.3mm]
\hspace*{-2mm} & \hspace*{-4mm} & \hspace*{-2mm} \Zm_{3,1}\rxv_1
\end{matrix}
& \hspace*{-4mm}\begin{matrix}
\Zm_{1,2}\rxv_2 \hspace*{-2mm} &\hspace*{-4mm}&\hspace*{-2mm} \\[-.3mm]
\hspace*{-2mm} & \hspace*{-2mm} \Zm_{2,2}\rxv_2 \hspace*{-2mm} \\[-.3mm]
\hspace*{-2mm} & \hspace*{-4mm} & \hspace*{-2mm} \Zm_{3,2}\rxv_2
\end{matrix}
\end{pmatrix} }_{\text{\normalsize $\triangleq \Xim$}}
\underbrace{\begin{pmatrix}
\rs_{1,1} \\[-.5mm]
\rs_{2,1} \\[-.5mm]
\rs_{3,1} \\[-.5mm]
\rs_{1,2} \\[-.5mm]
\rs_{2,2} \\[-.5mm]
\rs_{3,2} 
\end{pmatrix} }_{\text{\normalsize $\triangleq \rsv$}} \rmv .
\vspace{.3mm}
\ee

We will first
upper-bound $h(\ryv\ist |\ist\rxv)$. It follows from \eqref{eq:modelsimp} that given $\rxv$,
$\ryv$ is conditionally Gaussian with 
\vspace{.3mm}
covariance matrix $(\rho/2) \ist \Xim\Xim^{\operatorname{H}} \rmv+\Iv_{12}$. 
Hence, 
$h(\ryv\ist |\ist\rxv) 
\!=\rmv \E_{\rxv}\big[ \rmv\log\rmv\big((\pi e)^{12}\,\absdet{(\rho/2) \ist \Xim\Xim^{\operatorname{H}}$\linebreak 
$+\,\, \Iv_{12}} \ist \big)\big]$.
By \cite[Theorem~1.3.20]{hojo85}, 
$\absdet{(\rho/2) \ist \Xim\Xim^{\operatorname{H}} \rmv+\Iv_{12}} 
= \absdet{(\rho/2) \ist \Xim^{\operatorname{H}}\Xim +\rmv \Iv_{6}}$. Furthermore, assuming 
$\rho \rmv>\! 1$ (note that we are only interested in $\rho\rightarrow \infty$), we have
$\absdet{(\rho/2) \ist \Xim^{\operatorname{H}}\Xim +\rmv \Iv_{6}} \leq \rho^6\absdet{(1/2) \ist \Xim^{\operatorname{H}}\Xim +\rmv \Iv_{6}}$.
Thus,
\begin{align*}
h(\ryv\ist |\ist\rxv)& \ist\leq\ist\ist \E_{\rxv}\big[\log\rmv\big((\pi e)^{12}\rho^6 \,\absdet{(1/2) \ist \Xim^{\operatorname{H}}\Xim +\rmv \Iv_{6}} \ist \big)\big] \\[.5mm]
& \ist=\ist\ist 6\log (\rho) \ist+\ist \E_{\rxv}\big[\log\ist\absdet{(1/2) \ist \Xim^{\operatorname{H}}\Xim \ist+ \Iv_{6}} \ist \big]+\ist \mathcal{O}(1)\,.
\end{align*}
Finally, using 
$\E_{\rxv}\big[\log\ist\absdet{(1/2) \ist \Xim^{\operatorname{H}}\Xim \ist+ \Iv_{6}}\big] \leq \log \E_{\rxv}\big[\absdet{(1/2) \ist \Xim^{\operatorname{H}}\Xim$\linebreak 
$+\, \Iv_{6}}\big]= \mathcal{O}(1)$ 
\cite[Theorem 17.1.1]{Cover91}, 
we 
\pagebreak 
obtain
\be
h(\ryv\ist |\ist\rxv)  \ist\leq\ist 6 \log(\rho)+\ist \mathcal{O}(1)\,. \label{eq:boundhygivenx}
\ee

Next, we will lower-bound $h(\ryv)$. Using \eqref{eq:modelsimp}, we obtain
\begin{align*} 
h(\ryv) & \geq\ist h\bigg(\rmv\sqrt{\frac{\rho}{2}}\ist\ryvb+\rnv \ist\bigg|\ist \rnv\rmv\rmv\bigg) = h\bigg(\rmv\sqrt{\frac{\rho}{2}}\ist\ryvb\rmv\bigg) \notag \\[1mm]
& = 12 \log(\rho) \ist+\ist h(\ryvb) \ist+\ist \mathcal{O}(1) \,. 
\end{align*}
In Section \ref{sec:proof}, we will show that $h(\ryvb) \rmv > -\infty$. Hence,
$h(\ryv)\geq$\linebreak 
$12 \log(\rho)+\mathcal{O}(1)$ (note that $h(\ryvb)$ does not depend on $\rho$).
Inserting this bound and \eqref{eq:boundhygivenx} into \eqref{eq:sepixy}, we conclude that
$I(\rxv \ist;\ryv) \geq 6 \log(\rho)+\mathcal{O}(1)$. 
With \eqref{eq:preloggauss}, this implies $\chi \geq 3/2 = \T\ist (1\rmv-\rmv 1/\L)$.

\section{Proof that $\ist h(\ryvb) \rmv > -\infty$} \label{sec:proof}   

According to \eqref{eq:ybarsimp},
$\ryvb$ is a function of $\rsv$ and $\rxv$. We will relate 
$h(\ryvb)$ to $h(\rsv, \rxv)$.
To equalize the dimensions---note that $\ryvb\in \IC^{12}$ and $( \rsv^{\trans} \, \rxv^{\trans})^{\trans}\!\in \IC^{14}$---we 
condition on $[\rxv_1{]}_1$ and $[\rxv_2{]}_2$, which results in $h(\ryvb)\geq h(\ryvb\ist |\ist [\rxv_1{]}_1, [\rxv_2{]}_2)$.
For easier notation, we set
$\rxv_{\sP}\triangleq ([\rxv_1{]}_1 \,\ist [\rxv_2{]}_2)^\trans$ and 
$\rxv_{\sD}\triangleq ([\rxv_1{]}_2 \,\ist [\rxv_1{]}_3\,\ist [\rxv_1{]}_4$\linebreak 
$[\rxv_2{]}_1 \,\ist [\rxv_2{]}_3 \,\ist [\rxv_2{]}_4)^\trans\rmv$.
One can think of $\rxv_{\sP}$ as pilot symbols and of $\rxv_{\sD}$ as data symbols. The above inequality then becomes
\be\label{eq:boundhycond}
h(\ryvb) \ist\geq\ist h(\ryvb\ist |\ist \rxv_{\sP})\,.
\ee
We conclude the proof by showing that $h(\ryvb\ist \big|\ist\rxv_{\sP})>-\infty$. This will be done in the following five steps:
%
%
(i) Relate $(\sv, \xv_{\sD})$ to 
$\yvb$ via  polynomial mappings $\phi_{\xv_{\sP}}$. 
(ii) Show that the Jacobian matrices $\Jm_{\phi_{\xv_{\sP}}}\!(\sv, \xv_{\sD})\!$ are nonsingular almost everywhere (a.e.)\ for almost all (a.a.)\ $\xv_{\sP}$.
(iii) Show
that the mappings $\phi_{\xv_{\sP}}\rmv$ are finite-to-one a.e.\ for a.a.\ $\xv_{\sP}$.
(iv) Apply a novel result on the change in differential entropy 
under a finite-to-one mapping to $h(\ryvb\ist \big|\ist \rxv_{\sP})$.
(v) Bound the terms resulting
from this change in differential entropy.

\subsubsection*{Step (i)} We consider the $\xv_{\sP}$-parametrized mappings
\be\label{eq:phi}
\phi_{\xv_{\sP}} \colon (\sv,\xv_{\sD}) \,\mapsto\, \yvb \ist=
\begin{pmatrix} s_{1,1} \Zm_{1,1}\xv_1 + s_{1,2} \Zm_{1,2}\xv_2 \\[.5mm]
s_{2,1} \Zm_{2,1}\xv_1 + s_{2,2} \Zm_{2,2}\xv_2 \\[.5mm]
s_{3,1} \Zm_{3,1}\xv_1 + s_{3,2} \Zm_{3,2}\xv_2  
\end{pmatrix}\rmv,
\ee
which map $\IC^{12}$ to itself. 
The Jacobian matrix of 
$\phi_{\xv_{\sP}}$ is 
\[
\Jm_{\phi_{\xv_{\sP}}} = 
\begin{pmatrix}
\Bm &  
\begin{matrix}
\hspace{-1mm}\Am_{1,1} & \hspace{-2mm}\Am_{1,2} \\[0mm]
\hspace{-1mm}\Am_{2,1} & \hspace{-2mm}\Am_{2,2} \\[0mm]
\hspace{-1mm}\Am_{3,1} & \hspace{-2mm}\Am_{3,2} 
\end{matrix}
\end{pmatrix}\rmv,
\]
where $\Bm$ was defined in \eqref{eq:ybarsimp} and
\begin{align*}
\Am_{r,1} &\triangleq 
\begin{pmatrix}
\hspace*{-2mm} 0 \hspace*{-2mm}  \\[.5mm]
 s_{r,1}[\Qm_{r,1}{]}_2\hspace*{-2mm}  \\[0mm]
& \hspace*{-2mm} s_{r,1}[\Qm_{r,1}{]}_3\hspace*{-2mm}  \\[0mm]
&& \hspace*{-2mm} s_{r,1}[\Qm_{r,1}{]}_4
\end{pmatrix}\rmv,\\[1mm]
\Am_{r,2}\ist &\triangleq 
\begin{pmatrix}
 s_{r,2}[\Qm_{r,2}{]}_1 \hspace*{-2mm} \\[-.8mm]
& \hspace*{-2mm} 0 \hspace*{-2mm} \\[0mm]
&\hspace*{-2mm} s_{r,2}[\Qm_{r,2}{]}_3 \hspace*{-2mm} \\[0mm]
&&\hspace*{-2mm} s_{r,2}[\Qm_{r,2}{]}_4
\end{pmatrix}\rmv.
\end{align*}
Note that we did not take derivatives with respect to $[\xv_1{]}_1$ and $[\xv_2{]}_2$, since these variables are treated as fixed 
parameters.



\subsubsection*{Step (ii)}
To show that 
$\Jm_{\phi_{\xv_{\sP}}}\!$ is nonsingular (i.e., $\absdet{\Jm_{\phi_{\xv_{\sP}}}\rmv} \!\not=\! 0$) a.e.\ for a.a.\ $\xv_{\sP}$ and a generic $\Zm$, 
we use
the approach of
\cite[Appendix C]{koridumohl12}. 
The determinant of $\Jm_{\phi_{\xv_{\sP}}}\!$ is
a polynomial $p(\Zm, \sv, \xv)$ (i.e., a polynomial in all the entries of $\Zm$, $\sv$, and $\xv$), which we will show 
to be nonzero at a specific point $(\tilde{\Zm}, \svt, \xvt)$. 
Fixing $\svt$ and $\xvt$, we can then conclude
that $p(\Zm, \svt, \xvt)$ (as a function of $\Zm$) does not vanish identically. 
Since a polynomial vanishes either identically or on a set of measure zero, we conclude that $p(\Zm, \svt, \xvt)\neq 0$ for a generic $\Zm$. 
Using the same argument, we conclude that, for a generic fixed $\Zm$, 
$p(\Zm, \sv, \xv)\neq 0$ a.e.\ (as a function of $(\sv, \xv)$). 
Hence, $\absdet{\Jm_{\phi_{\xv_{\sP}}}\rmv} \!\not=\! 0$ a.e.\ for a.a.\ $\xv_{\sP}$ and a generic $\Zm$. 

\begin{figure}[tb]
\vspace{1.3mm}
\subfigure[]{\label{fig:matrix1}
	\includegraphics[trim = 34mm 207mm 127mm 23mm, clip,width=3.8cm]{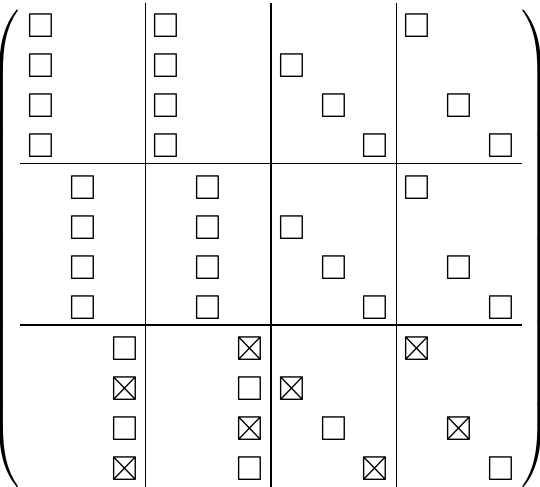}
}
\subfigure[]{\label{fig:matrix2}
	\includegraphics[trim = 42mm 223mm 135mm 23mm, clip,width=2.6cm]{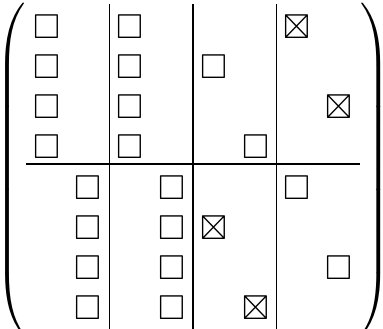}
}
\subfigure[]{\label{fig:matrix3}
	\includegraphics[trim = 51mm 240mm 143mm 23mm, clip,width=1.4cm]{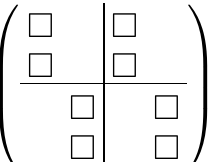}
}
\vspace{0mm}
\caption{Three matrices considered in Step (ii).
$\Box$ indicates a potentially nonzero entry; $\boxtimes$ indicates a potentially nonzero entry that is set to zero.
All the other entries are zero.}
\label{fig:matrices}
\vspace*{-5mm}\end{figure}

It remains to find the point $(\tilde{\Zm}, \svt, \xvt)$.
The matrix $\Jm_{\phi_{\xv_{\sP}}}\!$ has 
the form sketched in Fig.\ $\rmv$\ref{fig:matrix1}.
Setting $[\Qmt_{3,2}{]}_3 \!\rmv=\! [\Qmt_{3,1}{]}_4\!\rmv=\![\Qmt_{3,2}{]}_1$\linebreak 
$ \!=\! [\Qmt_{3,1}{]}_2 \!=\! 0$, 
the entries marked by $\boxtimes$ become zero.
Choosing $[\Qmt_{3,1}{]}_1$, $\rmv[\Qmt_{3,1}{]}_3$, $\rmv[\Qmt_{3,2}{]}_2$, $\rmv[\Qmt_{3,2}{]}_4$, 
$\rmv\tilde{s}_{3,1}$, $\rmv\tilde{s}_{3,2}$, $\rmv[\xvt_1{]}_1$, and $[\xvt_2{]}_2$ non\-zero 
and operating a Laplace expansion on the last four rows in Fig.\ $\rmv$\ref{fig:matrix1}, we see that the matrix in Fig.\ $\rmv$\ref{fig:matrix1} 
is non\-singular if the matrix in Fig.\ $\rmv$\ref{fig:matrix2} is nonsingular.
Setting $\tilde{s}_{1,2} \!=\! \tilde{s}_{2,1} \!=\! 0$, the entries
marked by $\boxtimes$ in Fig.\ $\rmv$\ref{fig:matrix2} become zero. 
By choosing $[\Qmt_{1,1}{]}_2$, $\rmv[\Qmt_{1,1}{]}_4$, $\rmv[\Qmt_{2,2}{]}_1$, $\rmv[\Qmt_{2,2}{]}_3$, $\rmv\tilde{s}_{1,1}$,
and $\rmv\tilde{s}_{2,2}$
nonzero and \nolinebreak 
oper\-ating \nolinebreak 
a Laplace expansion on the last 
four columns, it remains to show nonsingularity of the matrix in Fig.\ $\rmv$\ref{fig:matrix3}.
This can 
be achieved
by suitably choosing $[\Qmt_{1,1}{]}_1$, $\rmv[\Qmt_{1,1}{]}_3$, $\rmv[\Qmt_{1,2}{]}_1$, $\rmv[\Qmt_{1,2}{]}_3$, 
$\rmv[\Qmt_{2,1}{]}_2$, $\rmv[\Qmt_{2,1}{]}_4$, $\rmv[\Qmt_{2,2}{]}_2$, and $\rmv[\Qmt_{2,2}{]}_4$. 

\vspace{1mm}

\subsubsection*{Step (iii)}
By B\'ezout's theorem \cite[Proposition B.2.7]{VdE00},\linebreak 
$d$ multivariate polynomials of degree 
$k$ can have at most $k^d$ isolated common zeros.
Since the equation $\phi_{\xv_{\sP}}\rmv(\sv,\xv_{\sD})=\yvb$ can be reformulated as the system of polynomial equations $\phi_{\xv_{\sP}}\rmv(\sv,\xv_{\sD})-\yvb=\0v \in \IC^{12}\rmv$,
where each of the   
12 
polynomials is of degree two (see \eqref{eq:phi}), the points
$(\sv,\xv_{\sD})$ that are mapped by $\phi_{\xv_{\sP}}\!$ to the same $\yvb$ are the common zeros of 12 polynomials of degree two. 
Nonisolated common zeros of these polynomials can only exist in the set where 
$\Jm_{\phi_{\xv_{\sP}}}\!$ is singular. Hence, the set $\sM \triangleq \{(\sv,\xv_{\sD}) \rmv: \absdet{\Jm_{\phi_{\xv_{\sP}}}\rmv} \rmv\neq\rmv 0\}$ 
contains only isolated common zeros, whose number is upper-bounded by B\'ezout's theorem by $2^{12}\rmv$. It follows that the number of 
points $(\sv,\xv_{\sD}) \rmv\in\rmv \sM$ that are mapped by $\phi_{\xv_{\sP}}\!$ to the same $\yvb$ is upper-bounded by $2^{12}\rmv$, i.e.,  $\phi_{\xv_{\sP}}\big|_{\sM}$ is finite-to-one for a.a.\ $\xv_{\sP}$.
Because by Step (ii) the complement of the set $\sM$ has Lebesgue measure zero for a.a.\
$\xv_{\sP}$, the mapping $\phi_{\xv_{\sP}}\rmv$ is finite-to-one a.e.\ for a.a.\ $\xv_{\sP}$.


\subsubsection*{Step (iv)}
We will use the following novel 
result bounding the change in differential entropy 
under a finite-to-one mapping.
A proof is provided in the 
\vspace{1.5mm}
appendix.

\begin{lemma}\label{LEMchangeh}
Let $\ruv\in \IC^n$ be a random vector with continuous probability density function $f_{\ruv}$. Consider a 
continuously differentiable mapping $\vartheta\colon \IC^n \!\rightarrow \IC^n$ with Jacobian matrix $\Jm_{\vartheta}$. Let 
$\rvv \triangleq \vartheta(\ruv)$, and assume that the cardinality of the set $\vartheta^{-1}(\{\vv\})$ satisfies
$\abs{\vartheta^{-1}(\{\vv\})}\leq m < \infty$ a.e., for some 
$m\in \IN$ (i.e., $\vartheta$ is finite-to-one a.e.). Then:

(I) There exist disjoint measurable sets $\{\sU_k\}_{k\in [1:m]}$ such that $\vartheta\big|_{\sU_k}\!$ is one-to-one for each $k \rmv\in\rmv [1 \!:\rmv m]$ and 
\vspace{.4mm}
$\bigcup_{k\in [1:m]}\sU_k=\IC^n \rmv\setminus\rmv \sN$, where $\sN$ is a set of Lebesgue measure zero. 

(II) For any such sets \vspace{-1mm} $\{\sU_k\}_{k\in [1:m]}$,
\be
h(\rvv) \ist\geq\ist h(\ruv) +\rmv \int_{\IC^n} \!\rmv f_{\ruv}(\uv)\log (\absdet{\Jm_{\vartheta}(\uv)}^2) \,d\uv -H(\randk)\,, 
\label{eq:bounddiffe}
\vspace{-.5mm}
\ee
where $\randk$ is the discrete random variable that takes on the value $k$ when $\ruv\in\sU_k$ and $H$ denotes entropy.
\vspace{1.5mm}

\end{lemma}

Since by Step (iii) the mappings $\phi_{\xv_{\sP}}$ are finite-to-one a.e.\ for a.a.\ $\xv_{\sP}$, 
we can use Lemma~\ref{LEMchangeh} with $\ruv = (\rsv,\rxv_{\sD})$ and $\vartheta = \phi_{\xv_{\sP}}$. We thus 
\vspace{-1mm}
obtain
\begin{align*}
h(\ryvb\ist |\ist \rxv_{\sP}) &\,\geq\, h(\rsv,\rxv_{\sD}) 
  +\ist \E_{\rxv_{\sP}}\! \bigg[\rmv \int_{\IC^{12}} \! f_{\rsv,\rxv_{\sD}}(\sv,\xv_{\sD}) \\[-.5mm]
&\rule{14mm}{0mm} \times \log \rmv\big(\absdet{\Jm_{\phi_{\rxv_{\sP}}}\!(\sv,\xv_{\sD})}^2\big) \ist d(\sv,\xv_{\sD})  -H(\randk) \bigg]\ist. \\[-5.8mm]
\end{align*}


\subsubsection*{Step (v)}
The differential entropy $h(\rsv,\rxv_{\sD})$ 
is a finite constant, and the entropy $H(\randk)$ can be upper-bounded 
by the entropy of a uniformly distributed discrete random variable.
Hence, it remains to bound
\begin{align}
&\hspace{-1mm}\E_{\rxv_{\sP}} \!\bigg[\int_{\IC^{12}} \! f_{\rsv,\rxv_{\sD}}(\sv,\xv_{\sD}) \ist\log \rmv\big(\absdet{\Jm_{\phi_{\rxv_{\sP}}}\!(\sv,\xv_{\sD})}^2\big) 
  \ist d(\sv,\xv_{\sD})\bigg] \notag\\
&\rule{11mm}{0mm}=
\int_{\IC^{14}} \! f_{\rsv,\rxv}(\sv,\xv) \ist\log \rmv\big(\absdet{\Jm_{\phi_{\xv_{\sP}}}\!(\sv,\xv_{\sD})}^2\big) \ist d(\sv,\xv)\,.
\label{eq:finlogdet}
\end{align}
In \cite[Appendix C]{koridumohl12}, it is shown that for an analytic function $g\colon \IC^n \to \IC$ that is not identically zero,
\[ 
\int_{\IC^n} \!\exp(-\norm{\xiv}^2)\log(\abs{g(\xiv)})\,d\xiv \ist> -\infty \,.
\]
Since $f_{\rsv,\rxv}$ is the probability density function of a standard multivariate Gaus\-sian random vector and $\det \ist(\Jm_{\phi_{\xv_\sP}}\!(\sv,\xv_{\sD}))$ 
is a complex polynomial that is not identically zero as shown in Step (ii), it follows 
that the integral in \eqref{eq:finlogdet} is finite. Hence, $h(\ryvb\ist|\ist \rxv_{\sP})>-\infty$.
With \eqref{eq:boundhycond}, this concludes the proof that $h(\ryvb) \rmv >\rmv -\infty$.

\vspace{-.5mm}

\section*{Appendix:\, Proof of Lemma \ref{LEMchangeh}}

\vspace{.3mm}

Part (I), the separation of $\IC^n$ into measurable subsets $\sU_k$, can be shown using Zorn's Lemma (for details see \cite[Lemma 8]{koridumohl12}). 
To establish part (II), i.e., the bound \eqref{eq:bounddiffe}, we first note that
\vspace{-.5mm}
\be\label{eq:rvvdelta}
h(\rvv) \ist\geq\ist h(\rvv \ist |\ist \randk) \ist=\rmv \sum_{k\in [1:m]} \!\rmv h(\rvv\ist |\ist \randk \!=\! k) \, p_k\,,
\vspace{-.5mm}
\ee
where $p_k \triangleq\ist \operatorname{Pr}\ist [\ruv \rmv\in\rmv \sU_k] = \int_{\sU_k} \rmv f_{\ruv}(\uv)\ist d\uv$.
We assume without loss of generality that $p_k \!\neq\! 0$ for $k \!\in\! [1\!:\!m]$ 
(if $p_k \!=\! 0$ for some $k$, we simply omit
the corresponding term 
in \eqref{eq:rvvdelta}). 
Since $\vartheta\big|_{\sU_k}\!$ is one-to-one, 
$h(\rvv\ist|\ist\randk \!=\! k)$ can be transformed using the transformation rule 
for one-to-one mappings \cite[Lemma 3]{moridu13}: 
\be\label{eq:rvvdeltak}
h(\rvv\ist|\ist\randk \!=\! k) \ist=\ist h(\ruv\ist|\ist\randk \!=\! k) +\rmv \int_{\IC^n} \!f_{\ruv|\randk=k}(\uv)\log (\absdet{\Jm_{\vartheta}(\uv)}^2) \ist d\uv \ist.
\ee
The conditional probability density function of $\ruv$ given $\randk \!=\! k$ is $f_{\ruv|\randk=k}(\uv)=\mathbbmss{1}_{\sU_k}\rmv(\uv) \ist f_{\ruv}(\uv)/p_k$. Thus,
$h(\ruv\ist|\ist\randk \!=\! k) = - \int_{\ist \sU_k} \!\rmv \big(f_{\ruv}(\uv)/p_k\big) \log\rmv\big(f_{\ruv}(\uv)/p_k \big) \ist d\uv$, and \eqref{eq:rvvdeltak} becomes
\begin{align*}
h(\rvv\ist|\ist\randk \!=\! k) 
& \ist=\ist \frac{1}{p_k} \bigg[ \rmv-\! \int_{\sU_k} \!f_{\ruv}(\uv)\log\rmv\bigg( \rmv\frac{f_{\ruv}(\uv)}{p_k}\rmv \bigg) \ist d\uv  \notag\\
& \rule{16mm}{0mm} +\int_{\sU_k} \!f_{\ruv}(\uv)\log (\absdet{\Jm_{\vartheta}(\uv)}^2) \,d\uv \bigg] \notag\\
& \ist=\ist \frac{1}{p_k} \bigg[ \rmv-\! \int_{\sU_k} \!f_{\ruv}(\uv)\log\rmv\big(f_{\ruv}(\uv) \big) \ist d\uv  \notag\\[.5mm]
& \rule{10mm}{0mm} +\int_{\sU_k} \!f_{\ruv}(\uv)\log (\absdet{\Jm_{\vartheta}(\uv)}^2) \,d\uv \bigg]  + \log (p_k) \ist .
\end{align*}
Inserting this expression into \eqref{eq:rvvdelta} and recalling that the sets $\sU_k$ are disjoint and 
$\bigcup_{k\in [1:m]}\sU_k=\IC^n \rmv\setminus\rmv \sN$,
we 
\vspace{-.5mm}
obtain
\begin{align*}
h(\rvv) & \ist\geq\ist\ist -\! \int_{\IC^n}\! f_{\ruv}(\uv) \log\rmv\big(f_{\ruv}(\uv)\big) \ist d\uv \notag \\[.5mm]
& \rule{7mm}{0mm} + \int_{\IC^n}\! f_{\ruv}(\uv)\log (\absdet{\Jm_{\vartheta}(\uv)}^2) \ist d\uv \ist+\! \sum_{k\in [1:m]} \!p_k \log (p_k) \notag \\[-.5mm]
& \ist=\ist h(\ruv) \ist+\rmv \int_{\IC^n}\! f_{\ruv}(\uv)\log (\absdet{\Jm_{\vartheta}(\uv)}^2) \ist d\uv \ist-\ist H(\randk)\,,\\[-5.2mm]
\end{align*}
which is \eqref{eq:bounddiffe}.


\vspace{-.5mm}

\bibliography{references}
\bibliographystyle{IEEEtran}
\end{document}